\documentclass[twocolumn]{aastex63} 

\newcommand{\dif}{\mathrm{d}}

\usepackage{savesym}
\savesymbol{tablenum}
\usepackage{xurl}
\usepackage{amsmath}
\usepackage{siunitx}

\defcitealias{millholland2019obliquity}{ML19}
\defcitealias{eggleton1998equilibrium}{EKH}
\newcommand{\cta}{\citetalias}

\received{December 21, 2022}
\accepted{\today}

\submitjournal{ApJ}

\shorttitle{Self-Consistent Spin, Tidal and Dynamical EoMs in \texttt{REBOUNDx}}
\shortauthors{Lu et al.}

\graphicspath{{./}{figures/}}

\begin{document}

\title{Self-Consistent Spin, Tidal and Dynamical Equations of Motion in the \texttt{REBOUNDx} Framework}

\author[0000-0003-0834-8645]{Tiger Lu}
\affiliation{Department of Astronomy, Yale University, 52 Hillhouse, New Haven, CT 06511, USA}

\author[0000-0003-1927-731X]{Hanno Rein}
\affiliation{Department of Physical and Environmental Sciences, University of Toronto at Scarborough, Toronto, Ontario M1C 1A4, Canada}
\affiliation{David A. Dunlap Department of Astronomy and Astrophysics, University of Toronto, Toronto, Ontario, M5S 3H4, Canada}

\author[0000-0002-9908-8705]{Daniel Tamayo}
\affiliation{Department of Physics, Harvey Mudd College, Claremont, CA 91711}

\author[0000-0002-1032-0783]{Sam Hadden}
\affiliation{Canadian Institute for Theoretical Astrophysics, 60 St George St Toronto, ON M5S 3H8, Canada}

\author[0000-0001-7362-3311]{Rosemary Mardling}
\affiliation{School of Physics and Astronomy, Monash University, Victoria 3800, Australia}

\author[0000-0003-3130-2282]{Sarah C. Millholland}
\affiliation{MIT Kavli Institute for Astrophysics and Space Research, Massachusetts Institute of Technology, Cambridge, MA 02139, USA}

\author[0000-0002-3253-2621]{Gregory Laughlin}
\affiliation{Department of Astronomy, Yale University, 52 Hillhouse, New Haven, CT 06511, USA}

\correspondingauthor{Tiger Lu}
\email{tiger.lu@yale.edu}

\begin{abstract}

We have introduced self-consistent spin, tidal and dynamical equations of motion into \texttt{REBOUNDx}, a library of additional effects for the popular $N$-body integrator \texttt{REBOUND}. The equations of motion used are derived from the constant time lag approximation to the equilibrium tide model of tidal friction. These effects will allow the study of a variety of systems where the full dynamical picture cannot be encapsulated by point particle dynamics. We provide several test cases and benchmark the code's performance against analytic predictions. The open-source code is available in the most recent release of \texttt{REBOUNDx}.

\end{abstract}


\section{Introduction}
\label{sec:intro}
For a myriad of interesting astrophysical systems - including but not limited to close-in binaries, hot Jupiters, ultra-short period planets (USPs), and resonant chains - the dynamics of a planet's spin axis yield crucial insight regarding the state of the system. In general the spin axis has profound implications for climate and habitability - the rotation rate and obliquity of a planet greatly influences climate stability via effects on heat flux and radiative balance \citep{spiegel2009habitable}. In the specific case of the systems enumerated above, \textit{spin-orbit coupling} cannot be ignored - the dynamics of the system are impacted by the dynamics of the spin vector, and vice versa. In these cases separately computing the dynamics of the system and the evolution of the spin axes is insufficient - a self-consistent framework is required to fully capture the dynamics of both the system and the spin axes.

The \textit{equilibrium tide} model of tidal friction was first described by \cite{darwin1879xiii}, and has been expanded upon by many authors \citep{alexander1973weak,hut1980stability,hut1981tidal,eggleton1998equilibrium,mardling2002calculating}. In the Darwin model, a star or planet possesses a tidal bulge which lags by a constant small time interval from the orientation it would have in the absence of dissipation. The equations of motion governing the spin axis of each body depend on the body's quadrupole moment and the tidal forces acting on the quadrupole deformation, both of which in turn depend on the magnitude and direction of the spin vector.

A plethora of results that draw on the equilibrium tide model have been reported over the past century and a half. Enumerating a few examples relevant to the present discussion, \cite{goldreich1966} constrained the internal dissipation of the solar system planets based on secular orbital observations. \cite{wu_2003} invoked von Zeipel-Lidov-Kozai (ZLK) cycles and tidal friction to explain the highly eccentric orbit of HD 80606b \citep{naef_2001}. \cite{fabrycky2007shrinking} investigated the evolution and orbital distributions of binary stars using the same mechanism of ZLK cycles and tidal friction, and showed it to greatly enhance the population of binary stars on short-period orbits. \cite{mardling_2007} showed that for co-planar multi-planet systems at tidal fixed points such as HD 209458b \citep{charbonneau_2000}, eccentricity measurements yield incredible insights into the planet's internal structure. This approach was also applied by \cite{batygin_2009} to the HAT-P-13 system \citep{bakos_2009}, and generalized to inclined systems by \cite{mardling_2010}. The above studies all utilized orbit-averaged secular expressions. While significant insight can indeed be drawn from this analytic approach, they are inherently less flexible than \textit{N}-body simulations, and the orbit-averaging disallows analysis of resonant scenarios. 

Hence, in recent years the development of \textit{N}-body codes which self-consistently consider the evolution of both the spin and dynamical evolution of the system using instantaneous tidal deformation forces has been a priority. To provide a few excellent examples of such codes: \cite{millholland2019obliquity} used an \textit{N}-body code to show that secular resonance-driven spin-orbit coupling arising during disk-driven migration is able to generate and maintain large obliquities in many of the exoplanets discovered during the course of the Kepler Mission \citep{borucki_2010}. The resulting obliquity-driven tidal dissipation provides an evolutionary mechanism that can explain the overabundance of planets just wide of mean-motion resonance. \cite{Bolmont_2015} have modified the hybrid symplectic integrator \texttt{Mercury} \citep{chambers_1999} to consistently track the spin evolution in their package \texttt{Mercury-T}. They used this code to draw insights regarding the habitability of the Kepler-62 system \citep{borucki_2013} and showed the two planets in the systems habitable zone are likely to differ greatly in both obliquity and spin rate, with natural consequences for their habitability. \cite{kreyche2021exploring} further expanded upon this framework with \texttt{SMERCURY-T}, providing a framework that can self-consistently track the orbit and spin evolution of bodies in a multi-planet system under tidal influences from all bodies.  \cite{chen2021grit} present their independent consistent symplectic integrator package as well, \texttt{GRIT}. They have applied it to the Trappist-1 system \citep{gillon2017} and demonstrated that the differences in transit-timing variations could reach up to a few minutes over decade-long measurement baselines, and that strong planetary perturbations could push the outer Trappist-1 planets out of synchronized states.

In this work we present our implementation of self-consistent equations of spin, tidal and dynamical equations of motion in the \texttt{REBOUNDx} framework. \texttt{REBOUND} \citep{rein2012rebound} is a widely adopted open-source \textit{N}-body integrator package. \texttt{REBOUNDx} \citep{tamayo2020reboundx} is an associated library of routines that permits the flexible addition of additional physics to \texttt{REBOUND} simulations. Examples of such additions include exponential growth and damping of orbits, radiation forces, and post-Newtonian corrections. While an implementation of equilibrium tide theory is included in \texttt{REBOUNDx} \citep{baronett2022stellar}, this specific prescription does not evolve the spin axes of the bodies and hence is valid only for cases where spin-orbit coupling is negligible. This work introduces self-consistent equations of motion which are aware of the structure of each particle into the \texttt{REBOUNDx} package. While similar codes (mentioned previously) exist our framework provides unique advantages by virtue of its inclusion in \texttt{REBOUNDx}, including a variety of other rigorously-tested effects already present. These new effects in conjunction with the existing framework will provide many avenues of exploration.

The structure of the paper is as follows: in Section \ref{sec:ett} we describe the equilibrium tide model we have implemented, the approximations used and the reasoning behind our choices. In Section \ref{sec:eoms} we explicitly describe the coupled equations of motion and additional physical parameters needed to describe systems in the equilibrium tide model. In Section \ref{sec:implementation} we detail how our code fits into the \texttt{REBOUNDx} framework - readers not interested in the details of the background physics of equilibrium tide theory can skip to this section. In Section \ref{sec:tests} we apply the code to a few interesting test cases, and verify its accuracy with analytic results.

\section{Equilibrium Tide Theory}
\label{sec:ett}
The \textit{equilibrium tide} model was first described by \cite{darwin1879xiii}. In this model, a perturbed body assumes the shape it would have in hydrostatic equilibrium with the time-varying gravitational potential of the system. \cite{darwin1879xiii} expanded the gravitational potential of the perturbing body as a sum of Legendre polynomials. In the presence of internal dissipation, the body assumes a tidal deformation that is slightly misaligned with the line of centers connecting the two bodies. Each component contribution is associate with a frequency-dependent phase lag $\epsilon_\nu$, where $\nu$ is the frequency at which that component is forced.

The precise frequency dependence of the phase lag components is a complex function of internal structure, and hence extremely difficult to constrain. This has necessitated further approximations, for which two primary schools of thought have emerged - the \textit{constant time lag} model (CTL) and \textit{constant phase lag} (CPL). These prescriptions differ in their treatment of the phase lag components. We will briefly describe each approach.

The CTL approach directly follows from \cite{darwin1879xiii} in the limit of a visco-elastic body. \cite{alexander1973weak} was the first to evaluate the  \cite{darwin1879xiii} framework in this regime: each tidal component $\epsilon_l$ is directly proportional to the $l$th forcing frequency. This is equivalent to a fixed time lag $\tau$ between the tidal bulge and the line of centers between the two bodies, and $\tau$ is the constant of proportionality relating $\epsilon_l$ and the relevant tidal forcing frequency. \cite{hut1981tidal} provided a novel re-derivation by approximating the tidal bulge as two point masses on the surface of the body, and through energy \& angular momentum conservation arguments present orbit-averaged expressions for the evolution of the orbital elements and spin rate. \citet[hereafter EKH]{eggleton1998equilibrium} expands upon the \cite{hut1981tidal} framework by considering the distortion of the shape of a fluid planet to quadrupole order, in the presence of its own rotation $\mathbf{\Omega}$ and a tidal perturber. They then assumed that the rate of loss of tidal energy is directly proportional to the square of the rate of change of the shape (in the rotating frame), with the constant of proportionality being a dissipation constant $\sigma$ intrinsic to the interior structure of each body. This dissipation constant is related to the time lag $\tau$ (also intrinsic to each body) via

\begin{equation}
    \tau = \frac{3 \sigma r^5}{4G} k_L,
    \label{eq:ctl}
\end{equation}
where $r$ is the body's radius and $k_L$ its tidal Love number (parameterizing degree of central concentration - typically denoted $k_2$, but $k_L$ here to reduce confusion with other subscripts\footnote{Not to be confused with the \textit{apsidal motion constant}, notably denoted $k$ in many of the cited works and is equal to half the tidal Love number}). Both the \cite{hut1981tidal} and \cta{eggleton1998equilibrium} prescriptions have the advantage of introducing no discontinuities for low tidal frequencies, and makes no assumptions regarding eccentricity. While the \cite{hut1981tidal} framework assumes low obliquity, \cta{eggleton1998equilibrium} is valid for any orientation of the spin axis. This is the approach we have implemented.

We briefly summarize the alternative school of thought, the CPL approach. The CPL approach parameterizes the tidal response via the specific dissipation function $Q$ defined in \cite{goldreich1963}. Today this quantity is commonly referred to as the tidal quality factor:

\begin{equation}
    Q^{-1} \equiv \frac{1}{2 \pi E_0}\oint \left(-\frac{\dif E}{\dif t}\right)\dif t,
\end{equation}
where $E_0$ is the peak energy stored in the orbit during a tidal cycle, and $\oint \frac{\dif E}{\dif t} \dif t$ is the energy dissipated over a complete cycle. Via analogy to the simple harmonic oscillator \citep{Macdonald_1964, greenberg2009frequency}, $Q$ may be related to a component of phase lag $\epsilon$ via

\begin{equation}
    Q^{-1} = \tan 2 \epsilon.
\end{equation}

In principle, $Q$ is a function of the frequency and amplitude of the tidal perturbation. However, laboratory and field experiments showed that for a variety of solid materials $Q$ varies weakly, if at all, with frequency \citep{knopoff_macdonald_1958, knopoff_64}. Therefore, in the CPL prescription introduced by \cite{goldreich1966}, all tidal components are misaligned with the line of centers by the same angle $\epsilon$. This approximation reproduces the behavior of Earth and solid planets well, and is the prescription that has been widely adopted by the exoplanet community. While appealing in its simplicity, the CPL model poorly represents fluid bodies (whose $Q$ varies directly with frequency, see \citealt{knopoff_64}) and is only accurate to first order in eccentricity \citep{goldreich1963}. In addition, this model results in discontinuities when the tidal forcing frequency is close to zero \citep{heller2011tidal}.

Other excellent reviews contrasting the CTL and CPL approaches and the merits/drawbacks of either are \cite{greenberg2009frequency}, \cite{leconte2010tidal}, \cite{mardling_kitp} and \cite{heller2011tidal}. Given the previously listed disadvantages of the CPL model, we implement the CTL prescription of \cta{eggleton1998equilibrium} in this work. As the CPL prescription and parameterization of a $Q$ intrinsic to each planet has been adopted by the exoplanet community at large, for ease of use and intuition it is tempting to utilize some sort of hybrid approach. For instance, \cite{mardling2002calculating} present a fully self-consistent framework in the CTL regime, but implement $Q$ with the assumption that $Q$ can be expressed as some function of $\tau$. Such formulations are powerful in their accessibility. However, it is very important to note that these approaches are only valid in certain limiting cases - in general, there is no simple relation between $Q$ and $\tau$. In some cases of interest such a relation is possible \citep{leconte2010tidal}: for $Q\gg 1$ and a synchronized circular orbit, the eccentric annual tide with frequency $n$ (mean motion) dominates. In this case, we can write

\begin{equation}
    Q^{-1} \sim 2 \epsilon \sim 2 n \tau.
    \label{eq:q_approx}
\end{equation}

This is the assumption made in works including \cite{mardling2002calculating}, \cite{wu_2003} and \cite{millholland2019obliquity} to reconcile the tidal quality factor $Q$ with the CTL model. We emphasize again that use of the \cta{eggleton1998equilibrium} prescription is advantageous in that it makes no assumptions about the orbit of a tidal perturber (and hence is valid not just for satellite-primary tides, but for satellite-satellite tides as well). The user may appeal to this relation to set $\tau$ from a known value of $Q$, with the understanding that it is strictly valid only in the case of a synchronized circular orbit.

We recognize the contributions over the past decade by many authors \citep[among others]{ogilvie2004tidal, efroimsky2009tidal, ferraz2013tidal, correia2014deformation, storch2014viscoelastic, boue2019tidal, teyssandier2019formation, vick2019chaotic} in the development of more complex tidal formalisms. These tidal models more extensively account for the rheologies of the body, and may indeed yield more accurate and nuanced results than the equilibrium tide model. Implementing such models consistently into an \textit{N}-body integrator is certainly an avenue worth pursuing in the future. Given the present-day uncertainty around the precise interior compositions of exoplanets, the equilibrium tide model is elegant and powerful in its simplicity, and while exact quantitative details may differ it is more than sufficient to draw powerful qualitative insights.

\section{Equations of Motion}
\label{sec:eoms}
In this section, we explicitly describe the self-consistent spin, tidal and dynamical equations of motion of the \cta{eggleton1998equilibrium} we implement into \texttt{REBOUNDx}. These equations of motion represent an extension of the equilibrium tide theory already implemented in \texttt{REBOUNDx} by \cite{baronett2022stellar}, which assumed the framework of \cite{hut1981tidal} but assumed no evolution of the spin axes.

\begin{figure*}
    \centering
    \includegraphics[width=0.9\textwidth]{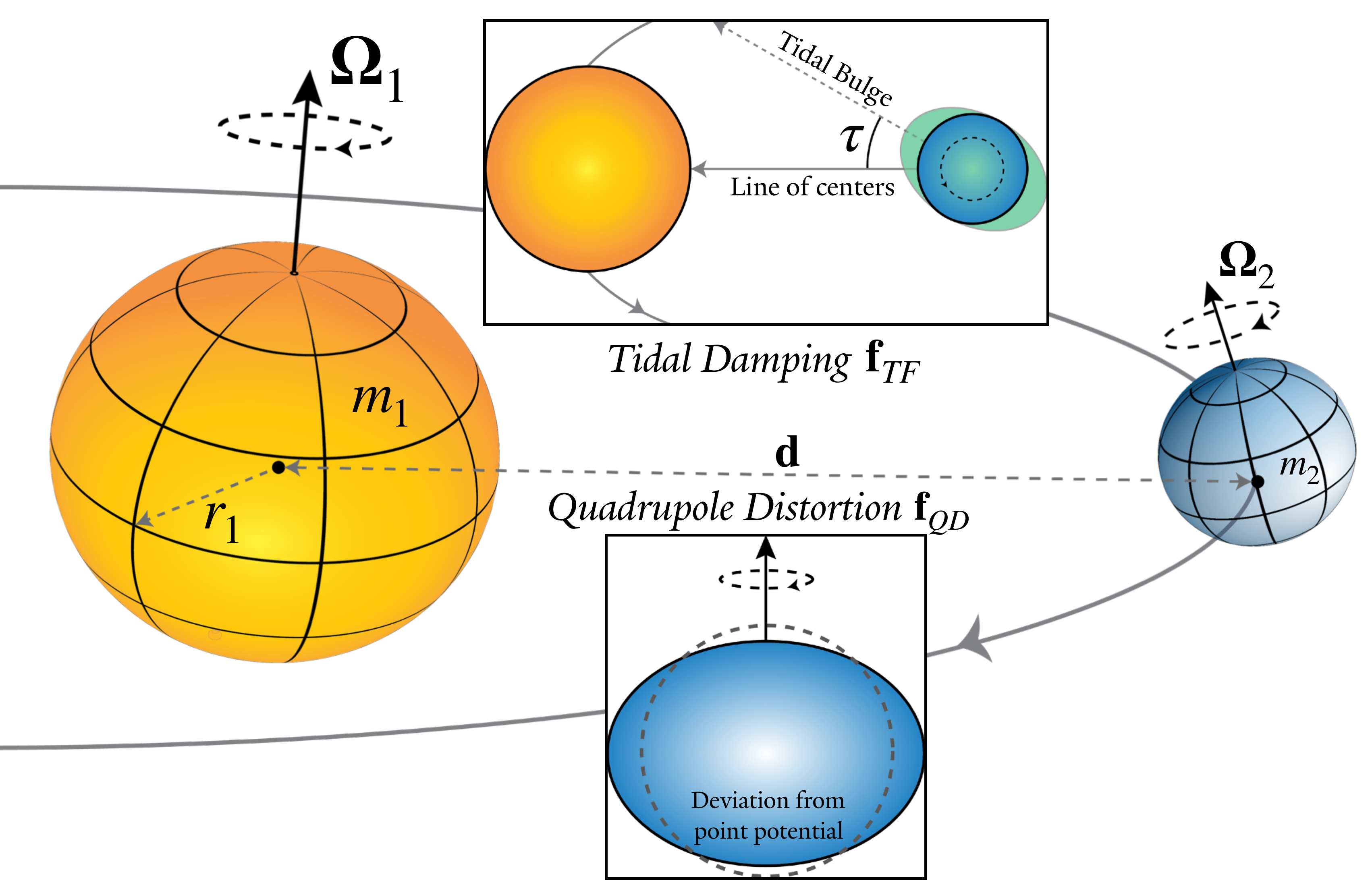}
    \caption{Schematic representation of the additional parameters, dynamical variables and forces needed to shift from a point particle framework to one in which the bodies are endowed with structure. Each body is parameterized by a mass $m$, a radius $r$, a spin vector $\mathbf{\Omega}$, a Love number $k_L$, and a dissipation constant $\sigma$ (directly related to the time lag $\tau$ via Equation \protect{\ref{eq:ctl}}), with the bodies separated by a separation vector $\mathbf{d}$. The additional forces to be considered are the quadrupole distortion force $\mathbf{f}_{\text{QD}}$ due to each body shape's deviation from a perfect sphere, and the tidal damping force $\mathbf{f}_{\text{TF}}$ arising from the dissipation of energy via tidal friction. Not pictured is the planet's obliquity $\theta$, defined as the angle between the spin axis and its orbit normal.}
    \label{fig:forces}
\end{figure*}

First, we list the parameters (set and fixed at the beginning of a simulation) and dynamical variables (set and evolved over time) associated with each body. For point particle dynamics, the only parameter that is required is the mass $m$, while the necessary dynamical variable is the vector distance $\mathbf{d}$ between each pair of bodies. Additional parameters and dynamical variables are required to describe dynamics beyond point particles. The parameters are, for each body: a radius $r$, the fully dimensional moment of inertia\footnote{While not directly used in the code, the dimensionless moment of inertia $C \equiv I / mr^2$ is a useful quantity that will be referenced later in the paper.}$I$, the Love number $k_L$, and the dissipation constant $\sigma$. The additional dynamical variables are, for each body, the three components of the spin vector $\Omega_x, \Omega_y, \Omega_z$ (parameterizing both the magnitude and direction of rotation). The equation of motion describing the relative vector separation $\mathbf{d} \equiv \mathbf{d}_1 - \mathbf{d}_2$ between two bodies (denoted $1$ and $2$) is given

\begin{equation}
    \ddot{\textbf{d}} = \textbf{f}_\text{g} + \textbf{f}_{\text{QD}}^{(1,2)} + \textbf{f}_{\text{QD}}^{(2,1)} + \textbf{f}_{\text{TF}}^{(1,2)} + \textbf{f}_{\text{TF}}^{(2,1)}.
\end{equation}
Explicitly, these forces are the familiar point particle gravitational acceleration:

\begin{equation}
    \textbf{f}_\text{g} = -\frac{G (m_1 + m_2)}{d^3} \textbf{d}.
\end{equation}
The acceleration due to the quadrupole moment of body~1, accounting for both its spin distortion and tidal distortion produced by body~2:

\begin{equation}
    \begin{split}
        \textbf{f}_{\text{QD}}^{(1, 2)} =  r_1^5 k_{L,1}\left(1 + \frac{m_2}{m_1}\right) \cdot \bigg[& \frac{5(\mathbf{\Omega}_1 \cdot \mathbf{d})^2 \mathbf{d}}{2d^7} - \frac{\Omega_1^2 \mathbf{d}}{2d^5} \\
        & - \frac{(\mathbf{\Omega}_1 \cdot \mathbf{d}) \mathbf{\Omega}_1}{d^5} - \frac{6 G m_2 \mathbf{d}}{d^8}\bigg],
    \end{split}
\end{equation}
with an equivalent expression for the acceleration due to the quadrupole moment of body~2, $\textbf{f}_{QD}^{(2, 1)}$. Furthermore, the acceleration due to the tidal damping of body~1 is:

\begin{equation}
    \begin{split}
        \textbf{f}_{\text{TF}}^{(1, 2)} = & -\frac{9 \sigma_1 k_{L,1}^2 r_1^{10}}{2 d^{10}} \left(m_2 + \frac{m_2^2}{m_1}\right) \\
        & \cdot \left[3 \mathbf{d} (\mathbf{d} \cdot \dot{\mathbf{d}}) + (\mathbf{d} \times \dot{\mathbf{d}} - \mathbf{\Omega}_1 d^2) \times \mathbf{d}\right],
    \end{split}
\end{equation}
and again an equivalent expression for the tidal damping of body 2. The addition of dissipative tidal forces into a symplectic scheme is not a concern with the operator-splitting methods applied by \texttt{REBOUND} - 
for an in-depth analysis see \cite{tamayo2020reboundx}. For a schematic representation of the relevant forces, see Figure \ref{fig:forces}. 

The evolution of the spin vector of body $1$ can be derived via matching torques on the orbit and the extended body in concordance with conservation of angular momentum, and is described by the differential equation

\begin{equation}
    I_1 \dot{\mathbf{\Omega}}_1 = -\left(\frac{m_1m_2}{m_1 + m_2}\right)\textbf{d} \times \left(\textbf{f}_{\text{QD}}^{(1, 2)} + \textbf{f}_{\text{TF}}^{(1, 2)}\right),
\end{equation}

with an equivalent expression for the evolution of body 2's spin vector. This framework may be extended to any number of bodies. We note here one subtlety that is not considered - each body reacts only in response to the tides it itself raises. For example, in a three-body system in which all three bodies are endowed with structure, body~2 reacts not only to the tides it raises on body~1 (which our code does consider), but also to the tides raised on body~1 by body~3. While these effects are expected to be minor for most systems, further work must certainly be done for a more complete picture that includes these non-pairwise accelerations.

\section{Implementation}
\label{sec:implementation}

In this section we describe the implementation of the previously described self-consistent spin, tidal and dynamical equations of motion in \texttt{REBOUNDx}\footnote{Detailed documentation is available at \href{https://reboundx.readthedocs.io/}{reboundx.readthedocs.io}.}.

\subsection{Spin, Structure and Tidal Parameters}
\label{subsec:params}
To turn on the additional forces for a given body, the following parameters \& dynamical variables must be set to finite values: the radius $r$, the Love Number $k_L$, the dissipation constant $\tau$, the moment of inertia $I$, and the spin vector components $\Omega_x, \Omega_y, \Omega_z$. 

\begin{itemize}
    \item If $k_L$ and all of $\Omega_x, \Omega_y, \Omega_z$ are set, the body will generate quadrupole distortion forces. If $\tau$ is also set, the body will generate tidal damping forces as well.
    
    \item If $I$ and all of $\Omega_x, \Omega_y, \Omega_z$ are set, the evolution of the body's spin axis will be tracked. Note that the evolution of the spin axis depends solely on $\textbf{f}_{\text{QD}} + \textbf{f}_{\text{TF}}$ - in other words, both $k_L$ and $\sigma$ should be set as well to observe any interesting dynamics of the spin vector; otherwise, the spin axis will remain stationary.
    
\end{itemize}

If none of these are set, the body will be treated as a point particle. Explicitly, it will raise tides on other bodies endowed with structure, and its own motion will be affected both by the associated quadrupole and tidal forces. However, it itself will generate neither. 

The framework of \cta{eggleton1998equilibrium} we have adopted parameterizes the magnitude of the tidal force via the dissipation constant $\sigma$ with units of $\text{mass}^{-1}\text{length}^{-2}\text{time}^{-1}$. $\sigma$ has a complex dependence on internal structure and we recognize that heuristic estimates of such a parameter will be unreliable and that few, if any, will have good intuition for reasonable values of $\sigma$. For this reason, leveraging that Equation \ref{eq:ctl} relating $\tau$ and $\sigma$ is always valid, the user-facing tidal parameter is chosen to be $\tau$. The selected value of $\tau$ is converted "under the hood" to the corresponding value of $\sigma$ for use in the equations of motion. It is worth noting that while direct measurements of $Q$ and $k_L$ exist for planets and satellites in the solar system \citep{lainey2009, lainey2016, lainey2017}, no such measurements exist for $\tau$. Hence, for many practical uses of the code the relation of Equation \ref{eq:q_approx} must be invoked to set $\tau$ from measured values. There are no built-in warnings for when this approximation is valid - it is left to the user's judgement to use this conversion as seen fit. 

\subsection{Dynamics}
\label{sec:dynamics}
The forces applied to each body are calculated in a pairwise manner. For a given pair of bodies, at each timestep the \texttt{REBOUNDx} parameters of each are checked to assess which forces it will generate. The relevant accelerations are then applied to each body, scaled to their relative masses. For example, in a system of two bodies $1$ and $2$, for the quadrupole distortion force generated by body $1$ $\mathbf{f}_{\text{QD}}^{(1, 2)}$: body $1$ will experience the acceleration $m_2 / (m_1 + m_2) \cdot \mathbf{f}_{\text{QD}}^{(1,2)}$, while body $2$ will experience the acceleration $m_1 / (m_1 + m_2) \cdot \left(-\mathbf{f}_{\text{QD}}^{(1,2)}\right)$.

The additional forces implemented are compatible with different integrators, and may also be used in conjunction with any of the other implemented effects in the \texttt{REBOUNDx} library. The package supports a mix of point particles and  bodies with internal structure in one simulation.

\subsection{Spin Axis Evolution}
The spin vectors of each body are tracked using \texttt{REBOUND}'s built-in coupled ODE structure.
This is a new feature that was added in \texttt{REBOUND} version 3.19. 
It allows the user to integrate any arbitrary set of ordinary differential equation structure in parallel with the main N-body integration\footnote{More information on this API can be found at \url{https://rebound.readthedocs.io/en/latest/ipython_examples/IntegratingArbitraryODEs/}.}.

Each set of ODEs can use the current dynamical state of the N-body simulation in their right-hand side equation.
Similarly, the current state of a user provided ODE can be used in calculating additional forces for the N-body particles. 

The user provided ODEs are integrated with an adaptive Gragg-Bulirsch-Stoer (\texttt{BS}) integrator and a default tolerance parameters of $10^{-5}$. 
If \texttt{BS} is also used for the integration of the N-body equations of motion, then everything is simply treated as one big coupled set of ordinary differential equations with one adaptive timestep. 

It is also possible to integrate arbitrary ODEs in conjunction with other \texttt{REBOUND} integrators such as \texttt{IAS15} and \texttt{WHFast}.
These integrators are typically more accurate and faster for integrations of planetary systems. 
In that case, only the user-defined ODEs are integrated with \texttt{BS} after a successful N-body integration step. 
\texttt{BS} still uses an adaptive timestep, but it also makes sure to synchronize its timesteps to that of the N-body integration.
This type of switching back and forth between N-body and user-provided ODEs will lead to an error. 
However, if the timescales involved in the user-defined ODEs are much longer than the timestep of the N-body integration then this will be a small error \citep{tamayo2020reboundx}.
This is typically the case for evolution of spin vectors.

To initialize the ODE structure associated with tracking the spin vectors, the user needs to call the \texttt{rebx\_spin\_initialize\_ode} function after all relevant \texttt{REBOUNDx} parameters have been set. 
This function sets up the ODE which tracks the spin vector evolution of every body with a valid moment of inertia $I$ and spin vector $\left[\Omega_x, \Omega_y, \Omega_z\right]$.
The spin vector \texttt{REBOUNDx} parameters $\Omega_x, \Omega_y, \Omega_z$ are updated before and after each ODE timestep - this means the user can pull the relevant real-time values of the spin axis by looking up the \texttt{REBOUNDx} parameters, without accessing the ODE framework itself.

At the moment, our implementation assumes there is a constant number of particles, and a constant number of particles for which we have to track the spin evolution. 

\section{Test Cases}
\label{sec:tests}

\subsection{Pseudo-Synchronization of Hot Jupiters}
\label{subsec:lec10}
A simple test of the code is a comparison with secular orbit-averaged analytic predictions. We use the analytic orbital evolution equations of \cite{leconte2010tidal}, which are the orbital evolution equations derived by \cite{hut1981tidal} extended to arbitrary obliquity and is equivalent to the \cta{eggleton1998equilibrium} framework. These equations describe a system of two mutually orbiting extended bodies, and describe the evolution of the semimajor axis, eccentricity, spin rate and obliquity of each body. The explicit expressions are given in Appendix \ref{appendix:lec10}.

\begin{figure*}
    \centering
    \includegraphics[width=0.95\textwidth]{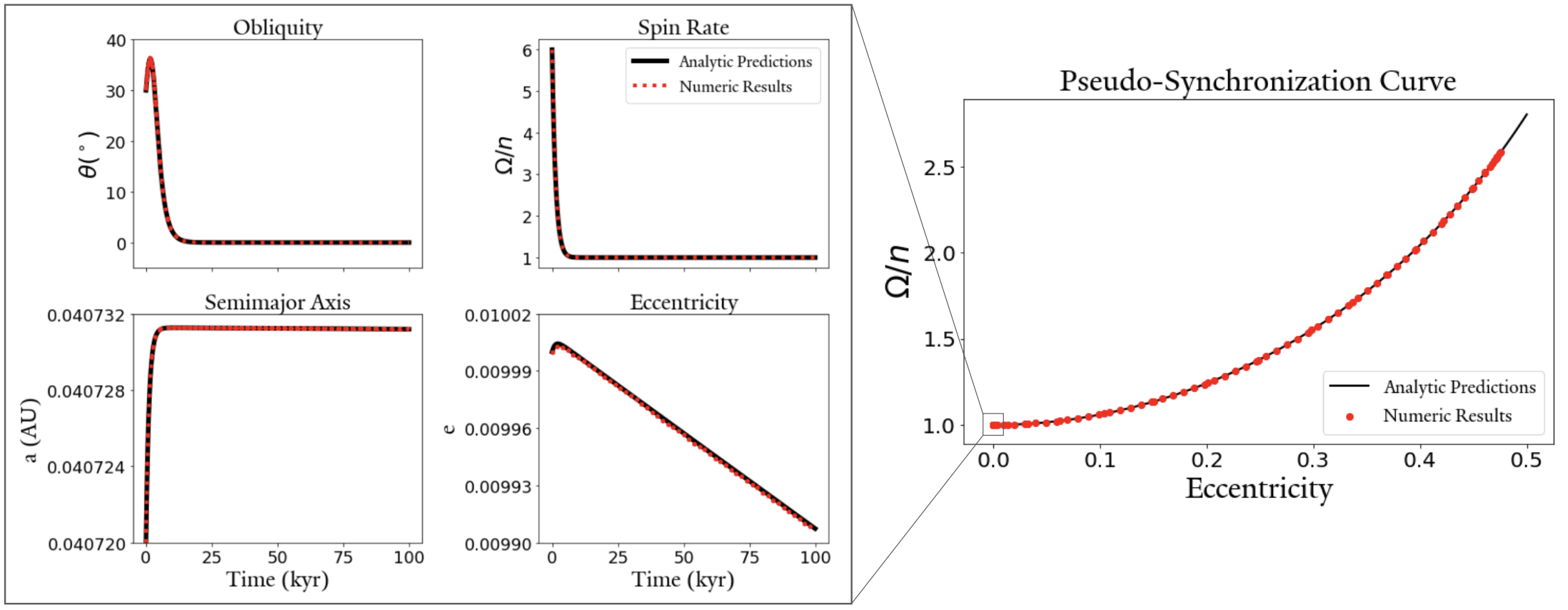}
    \caption{A comparison between the numerical results of our \texttt{REBOUNDx} simulations and analytic predictions using the framework of \protect{\cite{leconte2010tidal}}. The right subplot shows the end state of 80 numerical simulations of hot Jupiters initialized at varying eccentricities settling to their equilibrium pseudo-synchronous spin values, as well as the analytic prediction of these values given by Equation \protect{\ref{eq:pseudo_sync}}. Each simulation lands on the pseudo-synchronization curve and there is excellent agreement. The left subplot shows the evolution of the obliquity, spin rate, semimajor axis, and eccentricity of a specific hot Jupiter (corresponding to the lowest eccentricity case in the right subplot), where the black line represents the analytic predictions and the red dots represent our numeric results. The spin rate rapidly synchronizes and the obliquity is driven to zero. Again, there is excellent agreement between the analytic and numeric results.}
    \label{fig:lec10_comp}
\end{figure*}

We perform a basic test integration illustrating the spin-down and circularization of a generic hot Jupiter orbiting a Sun-like star with both the \texttt{REBOUNDx} numerical simulations and a numerical integration of the analytic equations of \cite{leconte2010tidal}, shown in the left subplot of Figure \ref{fig:lec10_comp}. The initial conditions of the simulation are: $m_* = 1 \text{M}_\odot, r_* = 1 \text{R}_\odot, m_p = 1 \text{M}_J, r_p = 1 \text{R}_J, a = 4.072 \times 10^{-2} \text{ AU}, e = 0.01, k_{L,*} = 0.07, \tau_* = 4.12 \times 10^{-4}\text{ s}, C_* = 0.07, \Omega_* = 27 \text{ days}, \theta_* = 0, k_{L,p} = 0.3, \tau_p = 4.12\text{ s}, C_p = 0.3,\Omega_p = 0.5\text{ days}, \theta_p = 30^\circ$. We used the symplectic \texttt{WHFast} integrator \citep{Rein_2015} with a timestep of one tenth the initial orbital period. This is the same test case used by \cite{millholland2019obliquity} to test the accuracy of their numeric code, and we see very close agreement between numeric and analytic results as they do.

One of the most notable predictions of the constant time lag model (noted upon both by \citealt{hut1981tidal} and \citealt{leconte2010tidal}) is rapid evolution towards a pseudo-synchronous state. More specifically, for typical hot Jupiter-like parameters, the planetary obliquity is quickly damped to zero while the spin rate evolves toward an equilibrium value given by

\begin{equation}
    \frac{\Omega_{eq}}{n} = \frac{1+\frac{15}{2}e^2+\frac{45}{8}e^4+\frac{5}{16}e^6}{\left(1+3e^2+\frac{3}{8}e^4\right)\big(1-e^2\big)^{3/2}}.
    \label{eq:pseudo_sync}
\end{equation}

To verify this prediction with our numeric code, we have run 80 additional simulations. The initial eccentricity of each of these additional simulations varied from $0.01$ to $0.8$, with all other parameters kept identical. Each simulation is advanced for $1\text{ Myr}$ and their final spin rates are reported on the right subplot of Figure~\ref{fig:lec10_comp}, as well as the analytically predicted equilibrium rotation values given by Equation \ref{eq:pseudo_sync}. We see excellent agreement and the "pseudo-synchronization" curve described in both \cite{hut1981tidal} and \cite{leconte2010tidal} is well reproduced. The earlier simple test integration corresponds to the lowest-eccentricity example.

\subsection{Obliquity-Driven Sculpting of Exoplanetary Systems}
\label{ss:ml19}
\begin{figure*}
    \centering
    \includegraphics[width=0.95\textwidth]{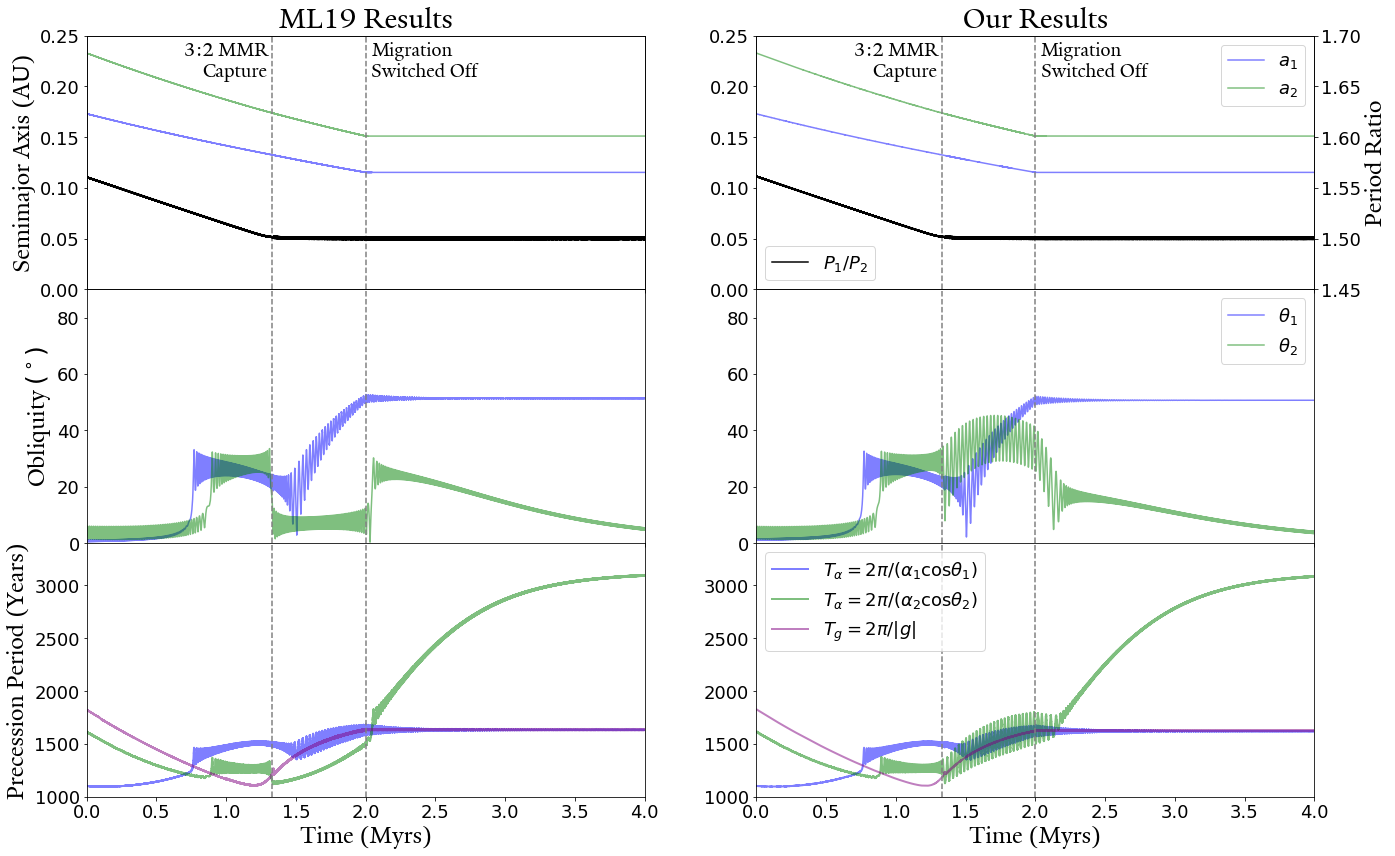}
    \caption{A comparison between \protect{\cta{millholland2019obliquity}}(left) and our results (right). The top subplot shows the semimajor axis evolution of both planets as they migrate inward and are caught into the 3:2 MMR. The blue points correspond to the inner planet, the green points to the outer planet, and the black line to their period ratios. The middle subplot shows the resulting obliquity evolution of the system. The dynamics of the system are identical - while the spin evolution differs the final state of the system is very similar. The bottom subplot shows the evolution of the relevant precessional frequencies of the system: the axial precession and nodal recession periods (see Appendix \protect{\ref{appendix:a}} for details).}
    \label{fig:sm19_comp}
\end{figure*}

An interesting result from the Kepler Mission \citep{borucki_2010} is the statistical excess of planet pairs just wide of first-order mean-motion resonances \citep{lissauer_2011, petrovich_2013,fabrycky_2014}. \citet[hereafter ML19]{millholland2019obliquity} postulate that obliquity tides could be a viable explanation; large axial tilts created by secular spin-orbit resonance spin-orbit coupling drain orbital energy to heat. Specifically, \cta{millholland2019obliquity} used an independent \textit{N}-body code to demonstrate that convergent migration and resonant interaction precipitated by capture into a mean-motion resonance is capable of generating and maintaining large obliquities over long periods, and argue that this mechanism is common in the compact, near-coplanar system typical of the \textit{Kepler} multis. In the test case used by \cta{millholland2019obliquity}, two planets are initialized around a star just wide of the 3:2 MMR. Both planets initially experience inward, convergent orbital migration, which is switched off after $2 \text{ Myr}$. See Figure \ref{fig:sm19_comp} for the dynamical evolution of the system: as the planets are captured into the 3:2 MMR, the inner planet is kicked to $>50^\circ$ obliquity, which is maintained indefinitely. While this particular simulation is associated with a specific set of parameters (see \cta{millholland2019obliquity} for details), no fine-tuning is done and the qualitative behavior of the system is independent of slight changes in initial parameters.

We reproduce the results of the \cta{millholland2019obliquity} simulation - Figure \ref{fig:sm19_comp} compares our two results. Our simulation is initialized in the same initial state as the \cta{millholland2019obliquity} simulation. The dynamics of the system use the \texttt{WHFast} integrator with a timestep of $0.159 \times 10^{-4}\text{ yrs}$ - this is roughly 0.1 times the inner planet's initial orbital period. The inward migration is modelled with the \texttt{modify\_orbits\_forces} implementation \citep{kostov2016, tamayo2020reboundx} in \texttt{REBOUNDx}: the $\tau_a$ parameter (describing the rate of migration) is set to $-5 \times 10^6 \text{ years}$ and $-4.54 \times 10^6 \text{ years}$ for the inner and outer planet, respectively (as in \cta{millholland2019obliquity}).

The qualitative end state behavior of the system - namely, the generation and maintenance of a high-obliquity state for the inner planet - is well reproduced. The quantitative differences in the evolution of the system are expected outcomes of the different implementations of the spin-orbit coupling framework used. While our framework is that of \cta{eggleton1998equilibrium}, \cta{millholland2019obliquity} uses that of \cite{mardling2002calculating} - these frameworks are equivalent with the assumption $Q^{-1} = 2 \epsilon = 2 n \tau$. \cta{millholland2019obliquity} uses an initial $Q = 10^4$ for both planets - this is equivalent to $\sigma_1 \sim 8.76 \times 10^{14}$ and $\sigma_2 \sim 1.36 \times 10^{15}$. In addition, $Q = 10^6$ is used for the central star - we assume this is the tidal quality factor associated with the forcing frequency of the inner planet (though qualitatively there is no difference for the particular problem), so $\sigma_* \sim 3.15 \times 10^3$ - all $\sigma$ values are given in the default \texttt{REBOUND} units\footnote{The default \texttt{REBOUND} units are $\textup{M}_\odot, \text{AU}, \text{yr}/2 \pi$ as measures of mass, distance, and time, respectively - this selection yields $G=1$.}. Otherwise, the physical parameters and initial conditions were identical to \cta{millholland2019obliquity}. In terms of system evolution, there are three differences between our framework and \cta{millholland2019obliquity}: \textbf{i)} \cta{millholland2019obliquity} assumes the planets do not raise tides on the star, or one another - the only tidal effects considered are the star raising tides on the planets. Our framework assumes each interacting pair raises tides upon one another. \textbf{ii)} \cta{millholland2019obliquity} does not employ the force splitting we have used (described in Section \ref{sec:dynamics}). \textbf{iii)} \cta{millholland2019obliquity} use a Bulirsch-Stoer integrator with a similar timestep to ours and an accuracy parameter of $10^{-13}$. These differences are expected to be minor and indeed do not affect the qualitative final state of the system significantly. A more in-depth dicussion regarding the evolutionary differences in the two codes may be found in Appendix \ref{appendix:a}.

\subsection{Exploring the ZLK effect}

It was shown by \cite{lidov_1962} and \cite{kozai_1962} that in a hierarchical three-body system characterized by a significant misalignment between the relative inclinations of the inner and outer orbits, there may be high-amplitude coupled oscillations in the eccentricity and inclination of the inner orbit. Commonly known as the Kozai-Lidov effect, the initial discovery of this mechanism by \cite{von_zeipel_1910} has been pointed out by \cite{ito_2019} who hence advocate referring to this effect as the von Zeipel-Lidov-Kozai (ZLK) effect, which we adopt here. This mechanism has since been greatly expanded upon and is well studied - for an in-depth review, see \cite{naoz_2016} and references within. Of particular relevance to this work is the effect of tidal friction when considered in conjunction with ZLK cycles. This was notably explored by \cite{wu_2003} regarding the orbit of HD 80606b and \cite{fabrycky2007shrinking} to explain the overabundance of short-period binary stars. 

The ZLK mechanism, coupled with tidal friction, has often been invoked to explain the unusual orbits of hot Jupiters and other close-in exoplanets. As the planet approaches periastron during the high-eccentricity epoch of a ZLK cycle, tidal dissipation becomes very significant and the planet's semi-major axis shrinks. Eventually, the ZLK cycles are damped out by tidal or general relativistic precessions \citep{einstein_1916}. The characteristic period of a ZLK cycle is given \citep{fabrycky2007shrinking}:

\begin{equation}
    \tau_{ZLK} = \frac{2 P_c^2}{3 \pi P_p} \frac{m_* + m_p + m_c}{m_c} (1 - e_c^2)^{3/2},
\end{equation}

where quantities subscripted with $*, c, p$ correspond to the central body, an outer perturber, and the planet respectively. Recall that the axial precession rate, $\alpha$, scales as $a_p^{-3}$. We define the ZKL frequency $g_{ZLK} \equiv 2 \pi / \tau_{ZKL}$ (in this section the traditional $g$ will be denoted $g_{orbit}$ for clarity). As the orbit of the planet shrinks $g_{ZLK} / \alpha$ will pass through unity, the criterion for resonance capture/kick. The effects of such a resonant crossing on the planet's obliquity\footnote{Obliquity here refers to the angle between the planet's orbit normal and its spin axis. This is not to be confused with the angle between the star's spin axis and the planet's orbit normal, another common designation known as \textit{stellar} obliquity.} have yet to be explored.

We explore the obliquity evolution of a generic Neptune-like planet in a binary star system experiencing ZKL oscillations. For concreteness, we initialize the initial conditions of the simulation to fiducial values. These are $m_* = m_c = 1.0 \text{ M}_\odot$, $r_* = \text{R}_\odot$, $k_{L,*} = 0.01$, $C_* = 0.07$, $m_p = 1 \text{ M}_N$, $r_p = \text{R}_N$, $a_p = 2 \text{ AU}$, $e_p = 0.01$, $k_{L,p} = 0.4$, $C_p = 0.25$, $a_c = 50 \text{ AU}$, $e_c = 0.7$, $i_c - i_p = 80^\circ$. In terms of the dissipation parameter, we adopt the approximation $Q = (2 \epsilon)^{-1} = (2 n \tau)^{-1}$ with initial fiducial values $Q_* = 10^6$ and $Q_p = 3 \times 10^5$. The spins of central star and planet are initialized with spin periods of $\Omega_* = 4.6 \text{ days}$ and $\Omega_p = 1 \text{ day}$, both with $0$ obliquity. The outer perturber is considered a point particle. We incorporate general relativistic precession via the \texttt{"gr\_full"} implementation in \texttt{REBOUNDx} \citep{newhall_1983, tamayo2020reboundx}. While the \texttt{WHFast} integrator was used for the two previous examples and in general has the fastest performance, in this case the high-eccentricity pericenter approaches inherent to a ZLK cycle would force \texttt{WHFast} to apply the worst-case timestep over the entire integration at great cost to  performance. For this reason, here we use the \texttt{IAS15} adaptive-timestep high-order integrator \citep{rein_ias15}. We use an initial timestep of $\dif t = 5 \times 10^{-2} \text{ years}$.

\begin{figure}
    \centering
    \includegraphics[width=0.48\textwidth]{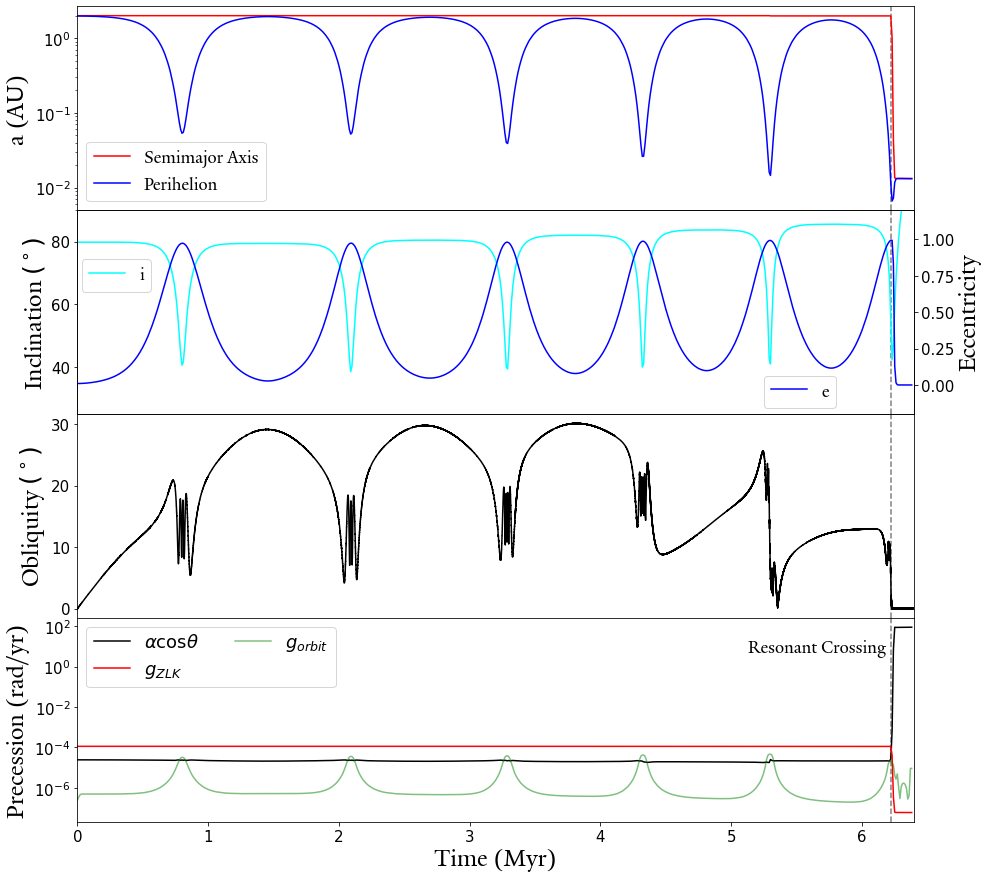}
    \caption{Results of the simulation demonstrating the ZLK mechanism with a fiducial Neptune-like planet. The top subplot shows the evolution of the planet's semimajor axis (red) and perihelion (blue). The second subplot shows the planet's eccentricity (blue) and mutual inclination between planet and perturber (aqua). The third subplot shows the obliquity of the planet, again defined as the angle between its orbit normal and spin axis. The bottom subplot shows the evolution of the three relevant precessional frequencies: $\alpha \cos \theta$, $g_{ZLK}$, and $g_{orbit}$. The gray dotted line spanning all four subplots denotes the point at which $g_{ZLK}$ crosses $\alpha \cos \theta$.}
    \label{fig:kozai}
\end{figure}

The results of this simulation are shown in Figure \ref{fig:kozai}, where we plot various quantities associated with the system's planet. Before $6 \text{ Myr}$, we see the standard ZLK oscillations. During the high-eccentricity epochs of each ZLK cycle, the planet experiences very quick nodal recession and the resulting crossings between $\alpha \cos \theta$ and $g_{orbit}$ results in several obliquity kicks in both directions. At $6.2 \text{ Myr}$, the semimajor axis rapidly shrinks, which motivates a rapid decrease in $g_{kozai}$ and increase in $\alpha \cos \theta$. When the two frequencies cross, the obliquity experiences a large kick downward as is damped to near zero, where it remains indefinitely. While the specific parameters of this simulation are given and changing these parameters does significantly alter the evolution of the system, this damping of obliquity at the crossing of $g_{kozai}$ and $\alpha \cos \theta$ is robust regardless of the specific parameters selected. We therefore conclude that planets which invoke ZLK cycles to explain their present orbits are expected to have negligible obliquity. This conclusion is particularly relevant given JWST observations of planets such as HD80606b expected in the near-future \citep{kataria_2021}, which may be able to constrain the obliquity of the planet.

\section{Summary}
In this work, we presented self-consistent spin, tidal and dynamical equations of motion integrated into the \texttt{REBOUNDx} framework \citep{tamayo2020reboundx}, as an improvement over the point-particle dynamics that \texttt{REBOUND} \citep{rein2012rebound} was previously restricted to. The equations of motion used are those derived in the constant time lag approximation of the equilibrium tide model by \cite{eggleton1998equilibrium}. The framework is set up such that these additional forces may be easily turned on or off, and such that a mix of bodies with structure and point particles may coexist in a \texttt{REBOUND} simulation. Extensive documentation and example Jupyter notebooks are available \href{https://reboundx.readthedocs.io/}{reboundx.readthedocs.io}.

We provide several interesting test cases and verify the results of the code against both analytic predictions \citep{leconte2010tidal} and previous work \citep{millholland2019obliquity}. We also apply the code to a fiducial system undergoing von Zeipel-Lidov-Kozai oscillations and show that these systems are expected to damp down to near zero obliquity at the conclusion of the oscillation period. These are just a few examples of the myriad applications of this framework. We anticipate our code to have wide-reaching applications to systems in resonant chains such as Trappist-1 \citep{gillon2017,luger2017,tamayo2017convergent,agol2021refining} and TOI-1136 \citep{dai2021}, mis/aligned systems \citep{rice2021soles, rice_2022}, ultra-short period planets \citep{Millholland_2020,dai2021}, and the general phase-space evolution of exoplanet obliquities \citep{su2022dynamics} to name a few. We hope this extension to \texttt{REBOUNDx} will provide a useful avenue in the study of exoplanet dynamics as a whole.

\acknowledgements
We thank Malena Rice, Alexander Heger, David Hernandez, Konstantin Batygin, Konstantin Gerbig and Tanvi Gupta for invaluable insight, discussion, and feedback. We also thank the anonymous reviewer for helpful comments that greatly improved the manuscript.

\software{\texttt{Jupyter} \citep{Kluyver2016jupyter}, \texttt{matplotlib} \citep{hunter_2007}, \texttt{numpy} \citep{harris2020array}, \texttt{pandas} \citep{mckinney2011pandas}, \texttt{REBOUND} \citep{rein2012rebound}, \texttt{REBOUNDx} \citep{tamayo2020reboundx}, \texttt{scipy} \citep{virtanen2020scipy}, \texttt{SLURM} \citep{yoo2003slurm}}

\appendix 
\section{Secular Orbit Evolution Equations}
\label{appendix:lec10}
The secular, orbit-averaged equations governing the evolution of the semimajor axis, eccentricity, spin rate and obliquity of each body for a system of two mutually orbiting extended bodies are \citep{leconte2010tidal}:

\begin{equation}
    \frac{\dif a}{\dif t} = \frac{4 a^2}{G m_1 m_2} \sum_{i=1}^2 K_i \left[ N(e) x_i \frac{\Omega_i}{n} - N_a(e)\right],
\end{equation}

\begin{equation}
    \frac{\dif e}{\dif t} = \frac{11 a e}{G m_1 m_2} \sum_{i=1}^2 K_i \left[ \omega_e(e) x_i \frac{\Omega_i}{n} - \frac{18}{11} N_e(e)\right],
\end{equation}

\begin{equation}
    \begin{split}
        \frac{\dif \Omega_i}{\dif t} = -\frac{K_i}{I_i n} \left[(1 + x_i^2) \omega(e) \frac{\Omega_i}{n} - 2 x_i N(e)\right],
    \end{split}
\end{equation}

\begin{equation}
    \begin{split}
        \frac{\dif \theta_i}{\dif t} = \frac{K_i \sin \theta_i}{I_i \Omega_i n} \left[\left(x_i - \eta_i\right) \omega(e) \frac{\Omega_i}{n} - 2 N(e)\right],
    \end{split}
\end{equation}
where $x_i = \cos \theta_i$, $N(e), N_a(e), N_e(e), \omega(e), \omega_e(e)$ are the functions of eccentricity:

\begin{equation}
    N(e) = \frac{1 + \frac{15}{2} e^2 + \frac{45}{8} e^4 + \frac{5}{16} e^6}{(1-e^2)^6},
\end{equation}

\begin{equation}
    N_a(e) = \frac{1 + \frac{31}{2} e^2 + \frac{255}{8} e^4 + \frac{185}{16} e^6 + \frac{25}{64} e^8}{(1-e^2)^{15/2}},
\end{equation}

\begin{equation}
    N_e(e) = \frac{1 + \frac{15}{4} e^2 + \frac{15}{8} e^4 + \frac{5}{64} e^6}{(1-e^2)^{13/2}},
\end{equation}

\begin{equation}
    \omega(e) = \frac{1 + 3e^2 + \frac{3}{8}e^4}{(1-e^2)^{9/2}},
\end{equation}

\begin{equation}
    \omega_e(e) = \frac{1 + \frac{3}{2}e^2 + \frac{1}{8}e^4}{(1-e^2)^5},
\end{equation}

$K_i$ is defined as\footnote{The $k$ used in \cite{leconte2010tidal} is the tidal Love number rather than the apsidal motion constant described by \cite{hut1981tidal}.}

\begin{equation}
    K_i = 3 k_{L,i} \tau_i \left(\frac{G m_i^2}{r_i}\right) \left(\frac{m_j}{m_i}\right)^2 \left(\frac{r_i}{a}\right)^6 n^2,
\end{equation}
and $\eta$ is the ratio of rotational to orbital angular momentum

\begin{equation}
    \eta_i = \frac{m_i + m_j}{m_im_j} \frac{I_i \Omega_i}{a^2 n \sqrt{1 - e^2}}.
\end{equation}

\section{Exploration of Numerical Differences}
\label{appendix:a}

In Section \ref{ss:ml19}, we benchmarked our results against the independent \textit{N}-body code of \cta{millholland2019obliquity}. We performed an integration on the same fiducial system they do, and while the outputs qualitatively match, the quantitative evolution significantly differs and merits exploration. Ultimately, these difference arise from sensitive dependence on the precise location of the spin axis. In this appendix we examine these differences in greater detail.

To understand these results it will be necessary to first provide a brief review of the process of secular spin-orbit resonance and the dynamics of the spin axis (for more in-depth reviews, see \citealt{ward_hamilton_2004}, \citealt{millholland_batygin_2019}, \citealt{su2020dynamics}, \citealt{lu_laughlin_2022}, as well as \cta{millholland2019obliquity} itself). We must first define two relevant dynamical quantities. The first is the \textit{axial precession} period $T_\alpha$, defined as the period at which the planet's spin axis precesses about its orbit normal due to torques from the host star and the planet's own rotation:

\begin{equation}
    T_\alpha \equiv \frac{2 \pi}{\alpha \cos \theta},
\end{equation}
where $\alpha$ is the precessional constant, which is given (in the absence of satellites) by \citep{ward_hamilton_2004, millholland2019obliquity}:

\begin{equation}
    \alpha = \frac{1}{2} \frac{M_*}{m} \left(\frac{r}{a}\right)^3 \frac{k_L}{C} \Omega.
\end{equation}

Where $M_*$ is the mass of the host star, $m$ is the planet mass, $r$ is the planet radius, and $a$ is the semimajor axis.The second relevant period is the \textit{nodal recession} period, the period at which the planet's orbit regresses about the invariant plane of the system. This period is given by $T_g = 2 \pi / |g|$, where the nodal recession rate $g$ is the rate of change of the longitude of ascending node of the planet's orbit, $\Omega_{node}$. The nodal recession rate arises from torques contributed by the other planets in the system. In a two-planet system, assuming negligible stellar oblateness the nodes of both planets regress at the same rate. For the comparison with \cta{millholland2019obliquity}, these periods are plotted on the bottom subplots of Figure \ref{fig:sm19_comp}.

The behavior of the spin vector is best understood in a frame of reference rotating along with the planet's orbital recession $g$, with $x$-axis aligned with the ascending node and $z$-axis along the orbit normal. To interpret our results, we first rotate our simulations into the invariant frame, with $z-$axis aligned with the total angular momentum of the system and $x$-axis aligned with the line of nodes. We transform from the invariant frame of the system to this frame via the transformation

\begin{equation}
    \hat{\mathbf{\Omega}}^* = A \hat{\mathbf{\Omega}},
\end{equation}

\noindent where $\hat{\mathbf{\Omega}}^*$ is the unit spin axis in the frame that rotates with planet, $\hat{\mathbf{\Omega}}$ is the unit spin axis in the invariant plane, and $A$ is the time-dependent rotation matrix

\begin{equation}
    A = \begin{bmatrix}
    \cos \Omega_{node} & \sin \Omega_{node} & 0 \\
    - \cos i \sin \Omega_{node} & \cos i \cos \Omega_{node} & \sin i \\
    \sin i \sin \Omega_{node} & - \sin i \cos \Omega_{node} & \cos i
    \end{bmatrix},
\end{equation}

In this frame, the trajectories of the unit spin vector trace out a family of parabolae - the exact landscape of spin-axis phase-space trajectories depends on the ratio $|g| / \alpha$. For the seminal work of \cite{colombo_1966}, this phase-space landscape of the spin axis is known as ``Colombo's Top'' (see also \citealt{peale_1974}, \citealt{ward_1975}, \citealt{henrard_1987} and \citealt{su2020dynamics}). In this frame of reference, the equilibrium states of the spin axis appear stationary. These equilibrium positions are known as ``Cassini States'', and depending on the exact value of $|g| / \alpha$ there are either two or four Cassini States. If $|g| / \alpha$ is below a critical value $(|g| / \alpha)_{crit}$, all four Cassini states exist. At $(|g| / \alpha)_{crit}$, Cassini States 1 and 4 merge and disappear - above $(|g| / \alpha)_{crit}$ only Cassini States 2 and 3 exist. The critical value is a function of inclination $i$:

\begin{equation}
    (|g|/\alpha)_{crit} = (\sin^{2/3}i + \cos^{2/3}i)^{-3/2} \sim 1 \text{ for small $i$.}
\end{equation}

\begin{figure*}
    \centering
    \includegraphics[width=0.8\textwidth]{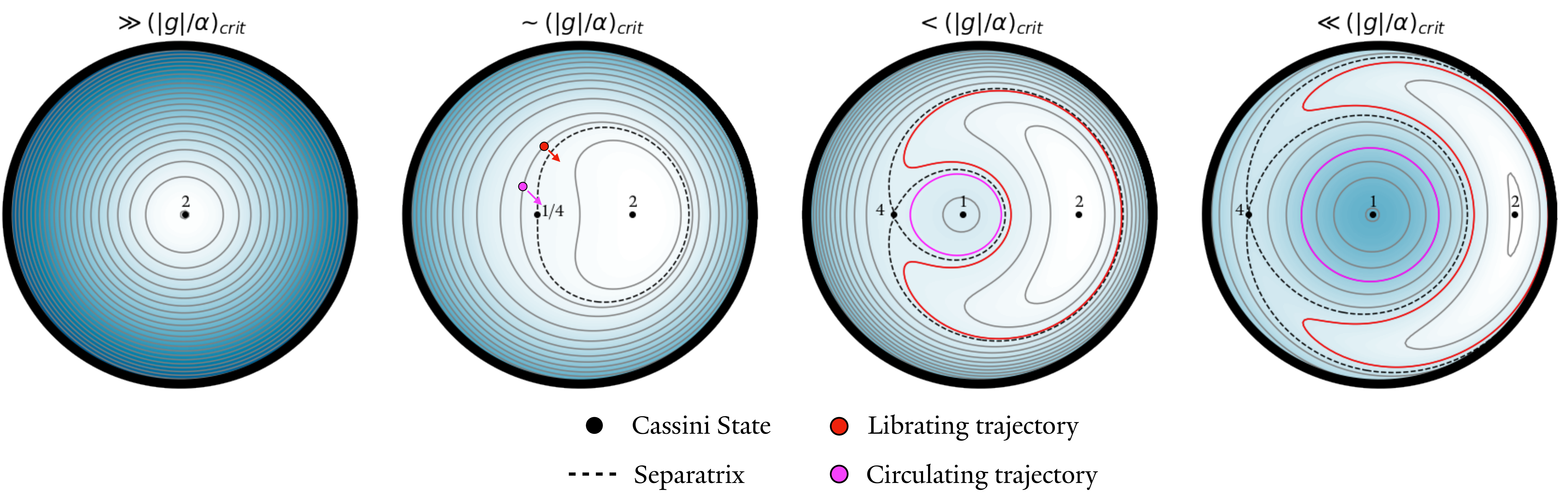}
    \caption{Schematic depiction of the topology of Colombo's top for various values of $|g| / \alpha$, along with the locations of the Cassini States. The red and pink dots/lines represent two initially similar spin states that, through a process of resonant kick and capture, ultimately end up on significantly different trajectories. This illustrates the sensitivity of the spin-axis evolution on the instantaneous phase at the time of resonant capture/kick.}
    \label{fig:spin_dyn}
\end{figure*}

Figure \ref{fig:spin_dyn} shows the topology of Colombo's Top for various values of of the ratio $|g| / \alpha$, as well as the location of the Cassini States (Cassini State 3 corresponds to a retrograde spin state and is not shown). Cassini States 1 and 2 are stable equilibria and adjacent trajectories will librate or circulate about them. Cassini State 4, on the other hand, lies on the separatrix and is unstable. 

Typically, two resonant mechanisms are invoked to explain high-obliquity states. Both are relevant to our current study. The first of these is resonant capture, and is motivated by the evolution of the topology of Colombo's Top as the ratio $|g| / \alpha$ decreases. If $|g| / \alpha$ approaches and crosses unity from above slowly enough to preserve the adiabatic criterion, a spin axis that begins on a low-obliquity state will follow Cassini State 2 as it evolves to high obliquity. The second of these mechanisms is a resonant kick. This may occur in a few ways, but for the case in interest this occurs upon rapid change of the ratio $|g| / \alpha$ such that the adiabatic criterion is not preserved. While resonant capture is characterized by movement of the trajectory itself, in a resonant kick the spin axis jumps from one trajectory to another.

Figure \ref{fig:spin_dyn} also overplots a demonstration of the sensitive dependence on the precise instantaneous location of the spin axis, invoking both resonant mechanisms. On the second subplot, two spin axis instances on the same trajectory are shown - the two instances will have very similar obliquities but different phase angles. If at the time shown both are given a small resonant kick, given their proximity to the separatrix they will be kicked onto trajectories in two different regimes - the red point is kicked onto a trajectory librating about Cassini State 2, while the pink point is kicked onto one circulating about Cassini State 1. As the system evolves as a consequence of the changing $|g| / \alpha$, the librating red trajectory is excited to higher and higher obliquity, while the pink trajectory does not change appreciably.

\begin{figure*}
    \centering
    \includegraphics[width=0.95\textwidth]{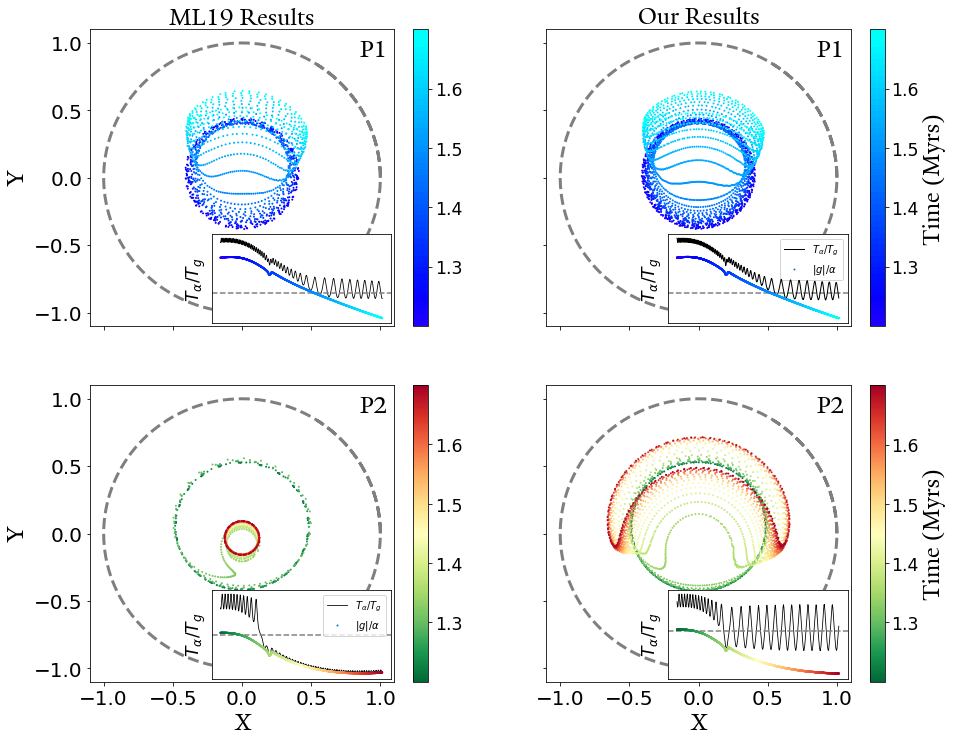}
    \caption{Scatterplots of the spin axis positions for the smaller time slice (from $1.2$ to $1.7$ Myr). The left column are results from \protect{\cta{millholland2019obliquity}}, while the right column are our results. The plots with the blue colorbar (first row) correspond to planet 1, while the plots with a green-red colorbar (second row) correspond to planet 2. Each subplot shows the position of the unit spin vector in the $XY$ plane (recall the $z$-axis is aligned with the planet's orbit normal) over time. Each subplot also shows an inset plot with three lines. The line with the same colorbar as the scatterplot represents the evolution of the ratio $|g| / \alpha$. The black line represents the ratio $T_\alpha / T_g$, which is equivalent to $|g| / \alpha \cos \theta$. The dashed grey line marks unity for clarity.}
    \label{fig:sm19_comp_zoom}
\end{figure*}

To more closely investigate the differences in our codes, Figure \ref{fig:sm19_comp_zoom} zooms in on a time slice of the simulation shown in Figure \ref{fig:sm19_comp} where the first significant deviation between the two results occurs $(1.2 \text{ Myr}$ to $1.7 \text{ Myr})$. The scatter plots show a polar view of the spin axis location, while the inset plots show the evolution of $|g| / \alpha$. Analysis of Figure \ref{fig:sm19_comp_zoom} clearly shows the mechanism described earlier in Figure \ref{fig:spin_dyn}: at roughly $1.3 \text{ Myr}$ a small kick is seen in the evolution of $|g| / \alpha$. The causes the spin axis to jump to a nearby trajectory - for planet 2, the slight difference in the spin axis position at the moment of the kick is enough to knock the spin axes onto what will become trajectories on diverging evolutions. The differences in evolution between our results and \cta{millholland2019obliquity} can be understood then as slight deviation in the physics from implementation differences resulting in significant differences due to the probabilistic nature of the resonant capture/kick process (for more review on the probabilistic nature of resonance, see \cite{henrard1982capture, saillenfest2021large}).

\bibliography{cite}{}
\bibliographystyle{aasjournal}

\end{document}